\documentstyle[12pt]{article}
\begin{document}
\def\esa{\begin{eqnarray*}}
\def\nsa{\end{eqnarray*}}
\addtolength{\baselineskip}{.3mm}
\thispagestyle{empty}
\begin{center}
{\large\bf Calabi--Yau Phases of the Superstring Vacuum}\\[13mm]
{\bf Yuri M. Malyuta}\\[3mm]
{\it Institute for Nuclear Research,
National Academy of Sciences of Ukraine\\
 252028 Kiev, Ukraine}\\[7mm]
{\bf Nikolay N. Aksenov}\\[3mm]
{\it  Space Research Institute,
  National Academy of Sciences of Ukraine\\
252022 Kiev,
Ukraine}\\
E-mail: aks@d310.icyb.kiev.ua\\[18mm]
\end{center}
\vspace{1cm}
\begin{center}
{\sc Abstract}
\end{center}
\noindent
It is shown that the model $X_{14}(7,3,2,1,1)$
has two Calabi--Yau phases.
\newpage

\section{Introduction}
It was discovered by Berglund, Katz and Klemm \cite{BKK}
that the model $X_{9}(3,2,2,1,1)$ has two Calabi--Yau phases.
Recently this type of models has been the focus of work in the context
of the heterotic/type II string duality \cite{KV,KLM}.

In this work we show that the model $X_{14}(7,3,2,1,1)$
also has two Calabi--Yau phases.
\section{Phase I}
Following the prescriptions of the papers  \cite{HKTY,HKTY1}
we can derive the generators of the Mori cone,
the principal parts of the Picard--Fuchs
operators and the normalization.

The generators of the Mori cone are
\esa
l^{(1)}& =& (0;7,0,1,-4,-2,-2),\\
l^{(2)}& =& (-2;-3,1,0,2,1,1).
\nsa
The principal parts of the Picard--Fuchs operators are
\esa
L_1&=&(7\theta_1 - 3\theta_2)\theta_1 ,\\
L_2&=&(\theta_2 - 2\theta_1)^3.
\nsa
The normalization is $K^0_{111}=9$.

With the computer program
INSTANTON \cite{HKTY1} we can obtain
 the Yukawa couplings and the instanton numbers.
The Yukawa couplings are
\esa
K_{111}& =& 220\,q_1\,{{q_2}^2} - 440\,q_1\,{{q_2}^3} +
   1100\,q_1\,{{q_2}^4} + 2300\,{{q_1}^2}\,{{q_2}^4}  \\
&&  - 7040\,q_1\,{{q_2}^5} + 184400\,{{q_1}^2}\,{{q_2}^5} +
   62920\,q_1\,{{q_2}^6} - 742200\,{{q_1}^2}\,{{q_2}^6} \\
&&  + 6160\,{{q_1}^3}\,{{q_2}^6} - 668360\,q_1\,{{q_2}^7} +
   6582800\,{{q_1}^2}\,{{q_2}^7} +
   407048112\,{{q_1}^3}\,{{q_2}^7}  \\
&& +  7891400\,q_1\,{{q_2}^8} -
   75772500\,{{q_1}^2}\,{{q_2}^8} -
   100073600\,q_1\,{{q_2}^9}    + \ldots \ ,
\nsa
\esa
K_{112}& =& 440\,q_1\,{{q_2}^2} - 1320\,q_1\,{{q_2}^3} +
   4400\,q_1\,{{q_2}^4} + 4600\,{{q_1}^2}\,{{q_2}^4}  \\
&&  - 35200\,q_1\,{{q_2}^5} + 461000\,{{q_1}^2}\,{{q_2}^5} +
   377520\,q_1\,{{q_2}^6} - 2226600\,{{q_1}^2}\,{{q_2}^6} \\
&&  + 12320\,{{q_1}^3}\,{{q_2}^6} - 4678520\,q_1\,{{q_2}^7} +
   23039800\,{{q_1}^2}\,{{q_2}^7}
 + 949778928\,{{q_1}^3}\,{{q_2}^7}  \\
&& +  63131200\,q_1\,{{q_2}^8} -
   303090000\,{{q_1}^2}\,{{q_2}^8} -
   900662400\,q_1\,{{q_2}^9}  + \ldots \ ,
\nsa
\esa
K_{122}& =& 880\,q_1\,{{q_2}^2} - 3960\,q_1\,{{q_2}^3} +
   17600\,q_1\,{{q_2}^4} + 9200\,{{q_1}^2}\,{{q_2}^4}
 -  176000\,q_1\,{{q_2}^5}  \\
&& +1152500\,{{q_1}^2}\,{{q_2}^5} +
   2265120\,q_1\,{{q_2}^6}
 -  6679800\,{{q_1}^2}\,{{q_2}^6} +
   24640\,{{q_1}^3}\,{{q_2}^6}  \\
&& -  32749640\,q_1\,{{q_2}^7} +
   80639300\,{{q_1}^2}\,{{q_2}^7}
  + 2216150832\,{{q_1}^3}\,{{q_2}^7}  \\
&& +  505049600\,q_1\,{{q_2}^8} -
   1212360000\,{{q_1}^2}\,{{q_2}^8} -
   8105961600\,q_1\,{{q_2}^9} + \ldots \ ,
\nsa
\esa
K_{222}& = &3\,q_2 - 45\,{{q_2}^2} +
   1760\,q_1\,{{q_2}^2} + 732\,{{q_2}^3} -
   11880\,q_1\,{{q_2}^3} - 12333\,{{q_2}^4}  \\
&& +  70400\,q_1\,{{q_2}^4} + 18400\,{{q_1}^2}\,{{q_2}^4} +
   211878\,{{q_2}^5} - 880000\,q_1\,{{q_2}^5} +
   2881250\,{{q_1}^2}\,{{q_2}^5} \\
&&   - 3685140\,{{q_2}^6}
 +  13590720\,q_1\,{{q_2}^6}
 -  20039400\,{{q_1}^2}\,{{q_2}^6} +
   49280\,{{q_1}^3}\,{{q_2}^6} \\
&&   + 64639725\,{{q_2}^7}
  - 229247480\,q_1\,{{q_2}^7}
 +  282237550\,{{q_1}^2}\,{{q_2}^7}
 +  5171018608\,{{q_1}^3}\,{{q_2}^7}  \\
&& -  1140830253\,{{q_2}^8} + 4040396800\,q_1\,{{q_2}^8}
 -  4849440000\,{{q_1}^2}\,{{q_2}^8} +
   20228948103\,{{q_2}^9}  \\
&& -  72953654400\,q_1\,{{q_2}^9}
  - 360011938170\,{{q_2}^{10}} + \ldots \ .
\nsa
The instanton numbers are recorded in Table 1.
\begin{center}
Table 1\\*[2mm]
\begin{tabular}{|r|r|r|r|r|r|}\hline
$n_{j,k}$&$j=0$
&$j=1$&$j=2$
&$j=3$&$j=4$\\ \hline
$n_{j,0}$&$0       $&$0      $&$ 0      $&$0       $&$0  $\\
$n_{j,1}$&$3       $&$0      $&$ 0      $&$0       $&$0  $\\
$n_{j,2}$&$-6      $&$220    $&$ 0      $&$0       $&$0  $\\
$n_{j,3}$&$27      $&$-440   $&$ 0      $&$0       $&$0  $\\
$n_{j,4}$&$-192    $&$1100   $&$ 260    $&$0       $&$0  $\\
$n_{j,5}$&$1695    $&$-7040  $&$ 23050  $&$0       $&$0  $\\
$n_{j,6}$&$-17064  $&$62920  $&$-92720  $&$220     $&$0  $\\
$n_{j,7}$&$188454  $&$-668360$&$ 822850 $&$15075856$&$0  $\\
$n_{j,8}$&$-2228160$&$7891400$&$-9471700$&$4571160 $&$260$\\\hline
\end{tabular}
\end{center}
\section{Phase II}
The generators of the Mori cone are
\esa
l^{(1)}& =& (-4;1,2,1,0,0,0),\\
l^{(2)}& =& (-2;-3,1,0,2,1,1).
\nsa
The principal parts of the Picard--Fuchs operators are
\esa
L_1&=&(\theta_1 - 3\theta_2)\theta_1 ,\\
L_2&=&\theta_2^3.
\nsa
The normalization is $K^0_{111}=9$.

Application of INSTANTON \cite{HKTY1} gives the Yukawa
couplings and the instanton numbers.
The Yukawa couplings are
\esa
K_{111}& =& 220\,q_1 + 2300\,{{q_1}^2} + 6160\,{{q_1}^3} +
   18940\,{{q_1}^4} + 27720\,{{q_1}^5} +
   64400\,{{q_1}^6} + 75680\,{{q_1}^7}  \\
&& +  152060\,{{q_1}^8} + 166540\,{{q_1}^9} +
   289800\,{{q_1}^{10}} - 440\,q_1\,q_2 +
   184400\,{{q_1}^2}\,q_2 \\
&&   + 407048112\,{{q_1}^3}\,q_2
 +   41624475840\,{{q_1}^4}\,q_2
  + 1828147967000\,{{q_1}^5}\,q_2 \\
&&  + 48876855669360\,{{q_1}^6}\,q_2
 +  928244560989200\,{{q_1}^7}\,q_2 +
   13662697686429696\,{{q_1}^8}\,q_2 \\
&& +  164713342061738520\,{{q_1}^9}\,q_2
 +  1100\,q_1\,{{q_2}^2} - 742200\,{{q_1}^2}\,{{q_2}^2} +
   123421320\,{{q_1}^3}\,{{q_2}^2}  \\
&& -  92301136560\,{{q_1}^4}\,{{q_2}^2}
 +  10878624450000\,{{q_1}^5}\,{{q_2}^2} +
   29170004481428400\,{{q_1}^6}\,{{q_2}^2} \\
&& +  4982997178421290560\,{{q_1}^7}\,{{q_2}^2}
 +  400064510148214852800\,{{q_1}^8}\,{{q_2}^2} -
   7040\,q_1\,{{q_2}^3} \\
&&   + 6582800\,{{q_1}^2}\,{{q_2}^3}
 -  1562042240\,{{q_1}^3}\,{{q_2}^3}
 +  383441785600\,{{q_1}^4}\,{{q_2}^3}  \\
&& -   57413729228000\,{{q_1}^5}\,{{q_2}^3} +
   3167226854530640\,{{q_1}^6}\,{{q_2}^3}  \\
&& -  10002877995518645440\,{{q_1}^7}\,{{q_2}^3}
 +  62920\,q_1\,{{q_2}^4} - 75772500\,{{q_1}^2}\,{{q_2}^4} \\
&& +  23668987320\,{{q_1}^3}\,{{q_2}^4}
 -  4915958054200\,{{q_1}^4}\,{{q_2}^4}
 +  799844578900500\,{{q_1}^5}\,{{q_2}^4}  \\
&& -  79798035056700600\,{{q_1}^6}\,{{q_2}^4}
 -  668360\,q_1\,{{q_2}^5}
 +  986861200\,{{q_1}^2}\,{{q_2}^5} \\
&& -  383570510400\,{{q_1}^3}\,{{q_2}^5}
 +  85307909769280\,{{q_1}^4}\,{{q_2}^5}
 -  14459980360680440\,{{q_1}^5}\,{{q_2}^5}\\
&& +  7891400\,q_1\,{{q_2}^6}
 -  13811657600\,{{q_1}^2}\,{{q_2}^6}
 +  6429424311200\,{{q_1}^3}\,{{q_2}^6} - \\
&& -  1615989224562160\,{{q_1}^4}\,{{q_2}^6}
 -  100073600\,q_1\,{{q_2}^7}
 +  202613621200\,{{q_1}^2}\,{{q_2}^7}\\
&& -  109939230054720\,{{q_1}^3}\,{{q_2}^7} +
   1336128420\,q_1\,{{q_2}^8}  \\
&& -  3072464694200\,{{q_1}^2}\,{{q_2}^8} -
   18546499400\,q_1\,{{q_2}^9} + \ldots \ ,
\nsa
\esa
K_{112}& =& -440\,q_1\,q_2 + 92200\,{{q_1}^2}\,q_2
 +  135682704\,{{q_1}^3}\,q_2
 +  10406118960\,{{q_1}^4}\,q_2  \\
&& +  365629593400\,{{q_1}^5}\,q_2
 +  8146142611560\,{{q_1}^6}\,q_2
 +  132606365855600\,{{q_1}^7}\,q_2  \\
&& +  1707837210803712\,{{q_1}^8}\,q_2
 +  18301482451304280\,{{q_1}^9}\,q_2
 +  2200\,q_1\,{{q_2}^2} \\
&& - 742200\,{{q_1}^2}\,{{q_2}^2}
 +  82280880\,{{q_1}^3}\,{{q_2}^2}
 -  46150568280\,{{q_1}^4}\,{{q_2}^2} \\
&& +  4351449780000\,{{q_1}^5}\,{{q_2}^2}
 +  9723334827142800\,{{q_1}^6}\,{{q_2}^2}  \\
&& +  1423713479548940160\,{{q_1}^7}\,{{q_2}^2}
 +  100016127537053713200\,{{q_1}^8}\,{{q_2}^2}  \\
&& -  21120\,q_1\,{{q_2}^3} + 9874200\,{{q_1}^2}\,{{q_2}^3}
 -  1562042240\,{{q_1}^3}\,{{q_2}^3}  \\
&& +  287581339200\,{{q_1}^4}\,{{q_2}^3}
 -  34448237536800\,{{q_1}^5}\,{{q_2}^3}
 +  1583613427265320\,{{q_1}^6}\,{{q_2}^3} \\
&& -  4286947712365133760\,{{q_1}^7}\,{{q_2}^3}
 +  251680\,q_1\,{{q_2}^4}
 -  151545000\,{{q_1}^2}\,{{q_2}^4}  \\
&& +  31558649760\,{{q_1}^3}\,{{q_2}^4}
 -  4915958054200\,{{q_1}^4}\,{{q_2}^4}
 +  639875663120400\,{{q_1}^5}\,{{q_2}^4} \\
&& -  53198690037800400\,{{q_1}^6}\,{{q_2}^4}
 -  3341800\,q_1\,{{q_2}^5}
 +  2467153000\,{{q_1}^2}\,{{q_2}^5}  \\
&& -  639284184000\,{{q_1}^3}\,{{q_2}^5}
 +  106634887211600\,{{q_1}^4}\,{{q_2}^5}
 -  14459980360680440\,{{q_1}^5}\,{{q_2}^5} \\
&& +  47348400\,q_1\,{{q_2}^6}
 -  41434972800\,{{q_1}^2}\,{{q_2}^6}
 +  12858848622400\,{{q_1}^3}\,{{q_2}^6} \\
&& -  2423983836843240\,{{q_1}^4}\,{{q_2}^6}
 -  700515200\,q_1\,{{q_2}^7}
 +  709147674200\,{{q_1}^2}\,{{q_2}^7}  \\
&& -  256524870127680\,{{q_1}^3}\,{{q_2}^7}
 +  10689027360\,q_1\,{{q_2}^8}
 -  12289858776800\,{{q_1}^2}\,{{q_2}^8} \\
&& -  166918494600\,q_1\,{{q_2}^9} + \ldots \ ,
\nsa
\esa
K_{122}& =& -440\,q_1\,q_2 + 46100\,{{q_1}^2}\,q_2
 +  45227568\,{{q_1}^3}\,q_2 + 2601529740\,{{q_1}^4}\,q_2 \\
&& +  73125918680\,{{q_1}^5}\,q_2
 +  1357690435260\,{{q_1}^6}\,q_2
 +  18943766550800\,{{q_1}^7}\,q_2 \\
&& +  213479651350464\,{{q_1}^8}\,q_2
 +  2033498050144920\,{{q_1}^9}\,q_2
 +  4400\,q_1\,{{q_2}^2} \\
&& - 742200\,{{q_1}^2}\,{{q_2}^2}
 +  54853920\,{{q_1}^3}\,{{q_2}^2}
 -  23075284140\,{{q_1}^4}\,{{q_2}^2} \\
&& +  1740579912000\,{{q_1}^5}\,{{q_2}^2}
 +  3241111609047600\,{{q_1}^6}\,{{q_2}^2}\\
&& +  406775279871125760\,{{q_1}^7}\,{{q_2}^2}
 +  25004031884263428300\,{{q_1}^8}\,{{q_2}^2}
 -  63360\,q_1\,{{q_2}^3} \\
&& + 14811300\,{{q_1}^2}\,{{q_2}^3}
 -  1562042240\,{{q_1}^3}\,{{q_2}^3}
 +  215686004400\,{{q_1}^4}\,{{q_2}^3} \\
&& -  20668942522080\,{{q_1}^5}\,{{q_2}^3}
 +  791806713632660\,{{q_1}^6}\,{{q_2}^3} \\
&& -  1837263305299343040\,{{q_1}^7}\,{{q_2}^3}
 +  1006720\,q_1\,{{q_2}^4}
 -  303090000\,{{q_1}^2}\,{{q_2}^4} \\
&& +  42078199680\,{{q_1}^3}\,{{q_2}^4}
 -  4915958054200\,{{q_1}^4}\,{{q_2}^4}
 +  511900530496320\,{{q_1}^5}\,{{q_2}^4} \\
&& -  35465793358533600\,{{q_1}^6}\,{{q_2}^4}
 -  16709000\,q_1\,{{q_2}^5}
 +  6167882500\,{{q_1}^2}\,{{q_2}^5}\\
&& -  1065473640000\,{{q_1}^3}\,{{q_2}^5}
 +  133293609014500\,{{q_1}^4}\,{{q_2}^5}
 -  14459980360680440\,{{q_1}^5}\,{{q_2}^5} \\
&& +  284090400\,q_1\,{{q_2}^6}
 -  124304918400\,{{q_1}^2}\,{{q_2}^6}
 +  25717697244800\,{{q_1}^3}\,{{q_2}^6} \\
&& -  3635975755264860\,{{q_1}^4}\,{{q_2}^6}
 -  4903606400\,q_1\,{{q_2}^7}
 +  2482016859700\,{{q_1}^2}\,{{q_2}^7} \\
&& -  598558030297920\,{{q_1}^3}\,{{q_2}^7}
 +  85512218880\,q_1\,{{q_2}^8}
 -  49159435107200\,{{q_1}^2}\,{{q_2}^8} \\
&& -  1502266451400\,q_1\,{{q_2}^9} + \ldots \ ,
\nsa
\esa
K_{222}& =& 3\,q_2 - 440\,q_1\,q_2
 +  23050\,{{q_1}^2}\,q_2 + 15075856\,{{q_1}^3}\,q_2
 +  650382435\,{{q_1}^4}\,q_2 \\
&& +  14625183736\,{{q_1}^5}\,q_2
 +  226281739210\,{{q_1}^6}\,q_2
 +  2706252364400\,{{q_1}^7}\,q_2 \\
&& +  26684956418808\,{{q_1}^8}\,q_2
 +  225944227793880\,{{q_1}^9}\,q_2 - 45\,{{q_2}^2} \\
&& +  8800\,q_1\,{{q_2}^2} - 742200\,{{q_1}^2}\,{{q_2}^2}
 +  36569280\,{{q_1}^3}\,{{q_2}^2}
 -  11537642070\,{{q_1}^4}\,{{q_2}^2} \\
&& +  696231964800\,{{q_1}^5}\,{{q_2}^2}
 +  1080370536349200\,{{q_1}^6}\,{{q_2}^2} \\
&& +  116221508534607360\,{{q_1}^7}\,{{q_2}^2}
 +  6251007971065857075\,{{q_1}^8}\,{{q_2}^2}
 +  732\,{{q_2}^3} \\
&& - 190080\,q_1\,{{q_2}^3}
 +  22216950\,{{q_1}^2}\,{{q_2}^3}
 -  1562042240\,{{q_1}^3}\,{{q_2}^3} \\
&& +  161764503300\,{{q_1}^4}\,{{q_2}^3}
 -  12401365513248\,{{q_1}^5}\,{{q_2}^3}
 +  395903356816330\,{{q_1}^6}\,{{q_2}^3} \\
&& -  787398559414004160\,{{q_1}^7}\,{{q_2}^3}
 -  12333\,{{q_2}^4}
 + 4026880\,q_1\,{{q_2}^4}
 -  606180000\,{{q_1}^2}\,{{q_2}^4} \\
&& +  56104266240\,{{q_1}^3}\,{{q_2}^4}
 -  4915958054200\,{{q_1}^4}\,{{q_2}^4}
 +  409520424397056\,{{q_1}^5}\,{{q_2}^4} \\
&& -  23643862239022400\,{{q_1}^6}\,{{q_2}^4}
 +  211878\,{{q_2}^5}
 - 83545000\,q_1\,{{q_2}^5} \\
&& +  15419706250\,{{q_1}^2}\,{{q_2}^5}
 -  1775789400000\,{{q_1}^3}\,{{q_2}^5}
 +  166617011268125\,{{q_1}^4}\,{{q_2}^5} \\
&& -  14459980360680440\,{{q_1}^5}\,{{q_2}^5}
 -  3685140\,{{q_2}^6}
 + 1704542400\,q_1\,{{q_2}^6} \\
&& -  372914755200\,{{q_1}^2}\,{{q_2}^6}
 +  51435394489600\,{{q_1}^3}\,{{q_2}^6}
 -  5453963632897290\,{{q_1}^4}\,{{q_2}^6} \\
&& +  64639725\,{{q_2}^7}
 - 34325244800\,q_1\,{{q_2}^7}
 +  8687059008950\,{{q_1}^2}\,{{q_2}^7}  \\
&& -  1396635404028480\,{{q_1}^3}\,{{q_2}^7}
 -  1140830253\,{{q_2}^8}
 + 684097751040\,q_1\,{{q_2}^8} \\
&& -  196637740428800\,{{q_1}^2}\,{{q_2}^8}
 +  20228948103\,{{q_2}^9}
 -  13520398062600\,q_1\,{{q_2}^9}  \\
&& -  360011938170\,{{q_2}^{10}} + \ldots \ .
\nsa
The instanton numbers are recorded in Table 2.
\begin{center}
Table 2\\*[2mm]
\begin{tabular}{|r|r|r|r|r|r|}\hline
$n_{j,k}$&$j=0$
&$j=1$&$j=2$
&$j=3$&$j=4$\\ \hline
$n_{j,0}$&$0       $&$220       $&$ 260         $&$220           $&$
260$\\
$n_{j,1}$&$3       $&$-440      $&$ 23050       $&$15075856      $&$
650382435$\\
$n_{j,2}$&$-6      $&$1100      $&$-92720       $&$4571160       $&$
-1442208140$\\
$n_{j,3}$&$27      $&$-7040     $&$ 822850      $&$-57853400     $&$
5991277900$\\
$n_{j,4}$&$-192    $&$62920     $&$-9471700     $&$876629160     $&$
-76811833000$\\
$n_{j,5}$&$1695    $&$-668360   $&$123357650    $&$-14206315200  $&$
1332936090145$\\
$n_{j,6}$&$-17064  $&$7891400   $&$-1726456320  $&$238126826300  $&$
-25249831736640$\\
$n_{j,7}$&$188454  $&$-100073600$&$25326702650  $&$-4071823335360$&$
490864423233165$\\
$n_{j,8}$&$-2228160$&$1336128420$&$-384058094640$&$70536923244480$&$
-9603462282665900$\\\hline
    \end{tabular}
\end{center}
>From Tables 1 and 2 we have the relation
\esa
n_{j,k+2j}(\mbox{phase I})=n_{j,k}(\mbox{phase II}).
\nsa


\newpage
\begin{thebibliography}{99}
\bibitem{BKK}
P.~Berglund, S.~Katz and A.~Klemm,
Mirror symmetry and the moduli space for generic hypersurfaces
in toric varieties,
  Nucl.~Phys. {\bf B456} (1995) 153.
\bibitem{KV}
S.~Kachru and C.~Vafa,
Exact results for $N=2$ compactifications of heterotic strings,
  Nucl.~Phys. {\bf B450} (1995) 69.
\bibitem{KLM}
A.~Klemm, W.~Lerche and P.~Mayr,
K3--Fibrations and heterotic/type II string duality,
{\tt hep-th/9506112}.
\bibitem{HKTY}
S.~Hosono, A.~Klemm, S.~Theisen and S.T.~Yau,
Mirror symmetry, mirror map and applications to Calabi--Yau
hypersurfaces, Commun.~Math.~Phys. 167 (1995) 301.
\bibitem{HKTY1}
S.~Hosono, A.~Klemm, S.~Theisen and S.T.~Yau,
Mirror symmetry, mirror map and applications
to complete intersection Calabi--Yau spaces,
Nucl.~Phys. {\bf B433} (1995) 501.
\end{thebibliography}
\end{document}